\documentclass[twocolumn,showpacs,pre,floatfix,superscriptaddress]{revtex4}

\usepackage{color} 
\usepackage{tabularx} 
\usepackage{epsfig}
\usepackage{amsmath} 
\usepackage{amssymb} 
\usepackage{graphicx}
\usepackage{wasysym}

\begin{document}

\title{Universality in phase boundary slopes for spin glasses on self
dual lattices}

\author{Masayuki Ohzeki}
\affiliation{Department of Systems Science, Graduate School of Informatics,
Kyoto University, Yoshida-Honmachi, Sakyo-ku, Kyoto 606-8501, Japan}

\author{Creighton K.~Thomas}
\affiliation{Department of Physics and Astronomy, Texas A\&M University,
College Station, Texas 77843-4242, USA}

\author{Helmut G.~Katzgraber} 
\affiliation {Department of Physics and Astronomy, Texas A\&M University,
College Station, Texas 77843-4242, USA}
\affiliation {Theoretische Physik, ETH Zurich, CH-8093 Zurich, Switzerland}

\author{H.~Bombin}
\affiliation{Perimeter Institute for Theoretical Physics, Waterloo,
Ontario N2L 2Y5, Canada}

\author{M.~A.~Martin-Delgado}
\affiliation{Departamento de F{\'i}sica Te{\'o}rica I, Universidad 
Complutense, 28040 Madrid, Spain}

\date{\today}

\begin{abstract}

We study the effects of disorder on the slope of the
disorder--temperature phase boundary near the Onsager point ($T_c =
2.269\ldots$) in spin-glass models.  So far, studies have focused on
marginal or irrelevant cases of disorder.  Using duality arguments,
as well as exact Pfaffian techniques we reproduce these analytical
estimates. In addition, we obtain different estimates for spin-glass
models on hierarchical lattices where the effects of disorder are
relevant.  We show that the phase-boundary slope near the Onsager
point can be used to probe for the relevance of disorder effects.

\end{abstract}


\maketitle

\section{Introduction}
\label{sec:introduction}

The critical behavior of lattice spin systems can change drastically
when disorder is introduced. Harris \cite{harris:74} demonstrated
that whenever $2 - d\nu < 0$ ($d$ the space dimension, $\nu$ the
critical exponent of the correlation length) the universality class
of the system {\em without} disorder can change when disorder
is introduced \cite{comment:exceptions}.  The two-dimensional
Ising model \cite{ising:25}---one of the simplest yet nontrivial
classical spin systems---has $d\nu= 2$. When disorder is
added one obtains the two-dimensional Edwards-Anderson Ising
spin glass \cite{edwards:75,binder:86}. However, the model is
marginal in that the disorder only changes the singularity in
the specific heat from a logarithmic to a double-logarithmic
behavior \cite{dotsenko:81,dotsenko:83}. Different numerical
studies have addressed this problem on square lattices
\cite{mcmillan:84b,merz:02,picco:06,hasenbusch:08c}, however, a
detailed study of the gradual increase of the perturbation of the
system on non-Bravais lattices remains to be performed.

In this work we study the effects of a gradual increase of the
disorder for small disorder perturbations for the two-dimensional
Ising model on different lattice geometries and, in particular,
non-Bravais lattices.  We study the slope of the phase boundary of
the disorder--temperature phase diagram in the limit of vanishing
disorder as a means to detect if disorder is a relevant perturbation
or not.  

The phase-boundary slope of the two-dimensional bond-diluted Ising
model has been studied in detail by different research groups
\cite{rapaport:72,osawa:73,domany:79}.  In particular, Domany's
perturbative approach gives an ``exact'' result for the slope in
the case of the bond-diluted Ising model, as well as the Ising model
with bimodal disorder on square lattices \cite{domany:79,domany:78}.
The results stems from a simple perturbation technique combined with
duality arguments, as well as the replica method.  The effective
Hamiltonian is separated into two parts: a nonrandom replicated Ising
model and an operator representing the effects of the disorder that
corresponds to interactions among different replicas.  The analysis,
starting from this effective Hamiltonian, is validated when the
effects of the disorder are not relevant from a renormalization-group
perspective.  Note that this approach can be applied to any self-dual
lattice as long as one assumes that {\em the disorder does not affect
the nature of the critical behavior}.

Therefore, the value of the slope has certain universal properties
shared within the self-duality and the irrelevant disorder.
Despite nonanalyticities known to be present in disordered Ising
models \cite{griffiths:69}, the disorder seems to merely give rise to
a linear shift in the critical point (for small disorder), because most
relevant operators in the disordered part of the effective Hamiltonian
are marginal in two dimensions.  Therefore, a naive perturbation that
evaluates only the shift of the critical point without any further
consideration can be justified and would give an exact value of
the slope of the phase boundary near the critical temperature in the
zero-disorder limit. For the case of the two-dimensional Ising model on
the square lattice, this would be the Onsager point [$T_{c}^{\rm IM} =
2/\ln(1+\sqrt{2})\approx 2.26918\ldots$] \cite{kramers:41,onsager:44}.
However, it is unclear if this is still valid when the disorder is
a relevant perturbation.

In this work we compute the phase boundary slope for different
self-dual lattices using duality arguments \cite{ohzeki:08,ohzeki:09}.
We can therefore infer from the value of the phase boundary slope
if the disorder is a relevant perturbation or not: If the obtained
results do not agree with Domany's prediction, we can infer that the
disorder is a relevant perturbation. Finally, we compare these results
to the phase boundary slope of the two-dimensional Ising model with
bimodal disorder computed using exact Pfaffian methods \cite{thomas:10},
also validated via Monte Carlo data.

\section{Hierarchical lattices}
\label{sec:hierarchical}

Because considerably larger system sizes can be computed using
hierarchical lattices, we first analyze the phase-boundary slope on
self-dual hierarchical lattices using the duality approach introduced
in Ref.~\cite{ohzeki:08}. Due to the self-duality \cite{kramers:41}
of the hierarchical lattices, the critical point is the same as for
the two-dimensional Ising model on the square lattice. However the
relevance of the disorder can change depending on the geometry of
the hierarchical lattice \cite{efrat:01}.

\subsection{Review: Duality analysis for spin glasses}
\label{sec:duality}

We first outline the duality analysis expanded to spin glasses.
The goal is to study the phase boundary of the bond-diluted and
bimodal Ising model in two space dimensions given by the Hamiltonian
\begin{equation}
{\mathcal H} = -J\sum_{\langle ij\rangle }\tau _{ij}S_{i}S_{j}.
\label{eq:ham}
\end{equation}
Here $S_i$ represent Ising spins and the sum is over nearest neighbors.
$J$ defines the energy scale.
For convenience, the partition function is written as a function of
the coupling constant $K=\beta J$ ($\beta = 1/T$, $T$ the temperature):
\begin{equation}
Z(K) = \sum_{\{S_i\}} 
       \prod_{\langle ij \rangle} {\rm e}^{K\tau_{ij} S_i S_j},
\label{eq:pf}
\end{equation}
where the couplings $\tau _{ij}$ are given by
\begin{equation}
P(\tau_{ij}) = p\delta(\tau _{ij}-1) + (1-p)\delta(\tau_{ij}+1)
\end{equation}
for the two-dimensional Ising model with bimodal disorder and
\begin{equation}
P(\tau _{ij}) = p\delta(\tau _{ij}-1) + (1-p)\delta(\tau_{ij})
\end{equation}
for the bond-diluted version of the model. For $p = 1$ the standard
duality analysis can be applied \cite{kramers:41,wu:76} to obtain
the exact location of the Onsager point (without having to evaluate
the free energy) by relating two partition functions at different
temperatures, i.e., 
\begin{equation}
Z(K) = \lambda^{N_{B}}Z(K^{\ast}).  
\label{Z2}
\end{equation}
Here $N_{B}$ is the number of bonds and $\lambda =
[1+\exp(-2K)]/\sqrt{2}$ is a ratio of the local principal Boltzmann
factors when the edge spins are parallel, i.e., $x_{0}=\exp(K)$
and its dual $x_{0}^{\ast}=[\exp(K)+\exp(-K)]/\sqrt{2}$.  The dual
coupling constant is given by $K^{\ast}=-\ln(\tanh K)/2$.  Under the
assumption that there is a single transition temperature, we identify
the critical point by the condition $K = K^{\ast} \equiv K_{c}$, i.e.,
the Onsager point \cite{kramers:41,wu:76,onsager:44}.  In this case
the prefactor $\lambda$ then becomes unity.

However, when disorder is present, a replica approach has to be
included to determine the critical point. In this case the effective
partition function is given by
\begin{equation}
Z_n(\{x\}) = 
\left[
\sum_{\{S^{(\alpha)}_i\}}
\prod_{\alpha=1}^n\prod_{\langle ij \rangle}{\rm e}^{K\tau_{ij} 
S^{(\alpha)}_i S^{(\alpha)}_j}
\right]_{\rm av} ,
\end{equation}
where $n$ is the replica number, and $[\cdots]_{\rm av}$ represents
an average over the disorder.  A similar relation as in Eq.~(\ref{Z2})
can be derived, namely
\begin{equation}
Z_n(\{x\}) = \lambda_{n}^{N_{B}}
Z_n(\{x^{\ast}\}) .
\label{Z3}
\end{equation}
Instead of the relationship between low and high temperatures
as in the case without disorder we establish a duality relation
between the local Boltzmann factors $x_{k}$ and $x_{k}^{\ast}$
\cite{ohzeki:09,nishimori:02,maillard:03}, where $k$ denotes
the number of anti-parallel pairs in $n$-replicated bonds.
The important local principal Boltzmann factor is the one with
the edge spins parallel, i.e., $x_{0}=[\exp(nK\tau _{ij})]_{\rm
av}$; the dual one is $x_{0}^{\ast }=[(\exp(K\tau _{ij}) +
\exp(-K\tau_{ij}))/\sqrt{2}\}^{n}]_{\rm av}$ given by the $2$-component
Fourier transform in par with standard duality procedures \cite{wu:76}.
It follows that $\lambda_{n}=x_{0}^{\ast }/x_{0}$.

Because the traditional duality arguments to find a fixed point
in an arbitrary replica number give unphysical results, we
establish the hypothesis that the prefactor $\lambda_{n}$ should
be unity at the critical point also for the case of spin glasses
\cite{nishimori:02,maillard:03}.  This assumption is verified as
correct by comparing results for different critical points to results
from other methods.

\begin{figure}
\includegraphics[width=\columnwidth]{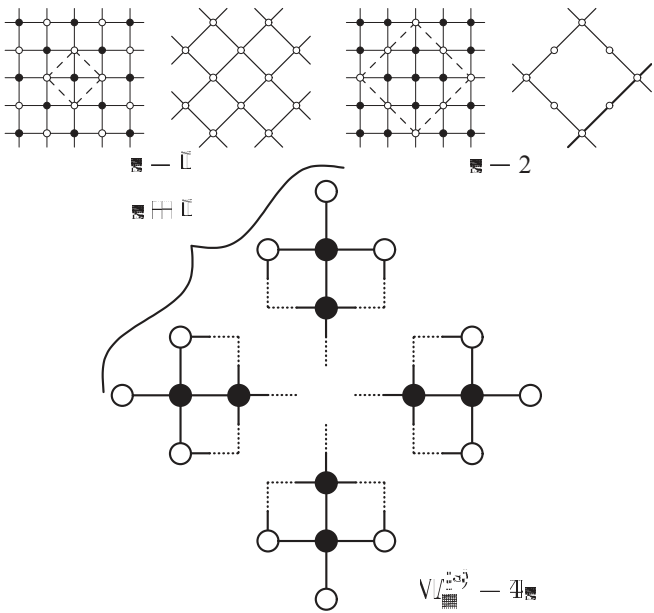}
\caption{
Clusters used in the summation over a subset of spins on a square
lattice for the improved duality approach. The full circles represent
the spins summed over, while the white circles are treated as fixed.
} 
\label{fig:clusters} 
\end{figure}

Recently, in Ref.~\cite{ohzeki:09} an improvement of this approach
has been proposed where one considers finite-size clusters to
regularly sum over a subsection of spins on the square lattice as a 
real-space renormalization after the ordinary duality transformation,
see Fig.~\ref{fig:clusters}. This accounts better for the effects
of the quenched randomness. Using this approach one obtains:
\begin{equation}
Z^{(s)}_n(\{x^{(s)}\}) = \left(\lambda _{n}^{(s)}\right)^{N_{B}/N_{B}^{(s)}}
Z^{(s)}_n(\{x^{\ast (s)}\}),  
\label{Z4}
\end{equation}
where $N_{B}^{(s)}$ is the number of bonds in the cluster, and $s$
denotes the size of the cluster.  The partition function is no
longer written by simple two-body interactions because of a regular
summation which we denote as $Z^{(s)}$ and $Z^{*(s)}$.  The local
Boltzmann factors $\{x\}$ and $\{x^*\}$ are modified by a regular
summation into $\{x^{(s)}\}$ and $\{x^{* (s)}\}$, which represents
these many-body interactions. The prefactor is also changed to
$\lambda^{(s)}_{n}=x^{*(s)}_{0}/x^{(s)}_{0}$, which is the ratio of
the renormalized principal Boltzmann factors with the spins on the
perimeter of the cluster fixed, as shown in Fig.~\ref{fig:clusters}
(fixed spins are represented by open circles).

A single equality---$\lambda _{n}^{(s)}=1$---then gives a more precise
location of the critical point in comparison to the standard duality
approach because the effects of the disorder are properly included.

\begin{figure}
\includegraphics[width=\columnwidth]{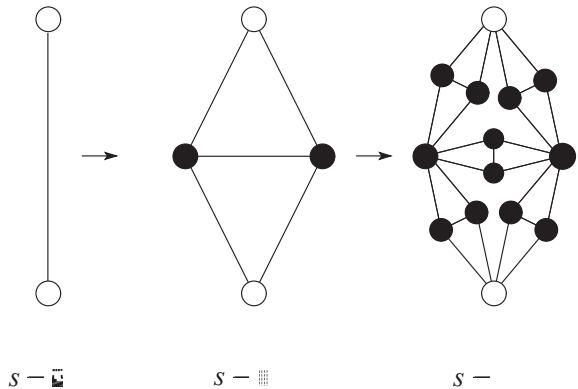}
\caption{
Self-dual hierarchical lattice. Full circles represent spins summed
over while the white circles are fixed.  The hierarchical lattice is
constructed by inserting a unit cell at each bond. The renormalization
corresponds exactly to the inverse process.
}
\label{fig:hlattice}
\end{figure}

The precision of critical points determined with the improved
duality approach increases with increasing cluster size, similar
to a renormalization-group analysis. When the studied lattice can
be built via a recursion---as in the case of hierarchical lattices
\cite{berker:79,griffiths:82}---the real-space renormalization group
analysis (the regular summation over a part of spins in each step
of the renormalization) can yield exact results asymptotically for
the partition function and related quantities \cite{ohzeki:08}.
In this case the cluster size $s$ represents the number of the
step in the construction instead of the cluster size, as shown in
Fig.~\ref{fig:hlattice}.

The prefactor $\lambda _{n}^{(s)}$ is explicitly given by the
ratio of $x_{0}^{(s)}=[Z_{\mathrm{cl}}^{(s)}(K)^{n}]_{\rm av}$ and
$x_{0}^{\ast(s)}=[Z_{\mathrm{cl}}^{\ast(s)}(K)^{n}]_{\rm av}$, where
$Z_{\mathrm{cl}}(K)$ represents the standard partition function as
in Eq. (\ref{eq:pf}) but limited to the cluster, and $Z^{\ast}(K)$
is its dual. The principal Boltzmann factors $x_{0}^{(s)}$ and
$x_{0}^{\ast(s)}$ can be regarded as the $n$-replicated partition
function on the finite-size cluster after a configurational (disorder)
average.  The quenched (replica) limit $n \to 0$ of $\lambda_{n}^{(s)}$
yields an equation that allows for the determination of the critical
points of spin glasses, namely
\begin{equation}
\left[ \log Z_{\mathrm{cl}}^{\ast (s)}(K)\right]_{\rm av} 
-
\left[ \log Z_{\mathrm{cl}}^{(s)}(K)\right]_{\rm av}
= 0 .  
\label{duality5}
\end{equation}

Starting from this equation, for instance, one can compute to high
precision the location of the multicritical point for the bimodal
($\pm J$) Ising model ($p_{c}=0.89082$) using the $s=2$ cluster on the
square lattice.  This also proves that there is no finite-temperature
spin-glass transition when the disorder distribution is symmetric
($p = 0.5$) on {\em any} self-dual lattice in two space dimensions
\cite{ohzeki:09,ohzeki:09b}.

\subsection{Computing phase-boundary slopes}
\label{sec:slope}

In the present study we estimate the value of the critical phase
boundary slope near the Onsager point for the bimodal ($\pm J$) Ising
model and the bond-diluted Ising model.  After that, we compare our
results with the standard perturbation result for the bimodal case
by Domany for the phase boundary slope $\zeta$, namely \cite{domany:79}
\begin{equation}
\zeta \equiv \frac{1}{T_{c}^{\rm IM}}
\left.\frac{\Delta T}{\Delta p}\right|_{p \to 0}
= \frac{1}{K_{c}\langle \sigma_{i}\sigma _{j}\rangle _{K_{c}}}
= 3.2091\ldots ,
\end{equation}
where $\langle \sigma _{i}\sigma _{j}\rangle _{K_{c}}$ is the
correlator between nearest-neighbor spins, which is the same
as the internal energy divided by the number of bonds.  $T_{\rm
c}^{\rm IM} = 2/\ln(1+\sqrt{2}) \approx 2.26918$ is the critical
temperature of the two-dimensional Ising model without disorder
\cite{kramers:41,onsager:44} shared within the self-dual hierarchical
lattices.  To compute the slope of the phase boundary close to the
Onsager point we expand Eq.~(\ref{duality5}) to first order for
$\Delta p\ll 1$
\begin{eqnarray}
&&\left( 1-N_{B}^{(s)}\Delta p\right) \left( \log Z_{0}^{\ast (s)}(K)-\log
Z_{0}^{(s)}(K)\right)   \notag \\
&&\quad +\Delta p\sum \left( \log Z_{1}^{\ast (s)}(K)-\log
Z_{1}^{(s)}(K)\right) =0.
\label{eq:lala}
\end{eqnarray}
Here $N_{B}^{(s)}$ denotes the number of bonds on the
finite-size cluster.  For the bimodal ($\pm J$) Ising model,
$Z_{0}$ and $Z_{0}^{\ast}$ are the partition functions without any
antiferromagnetic interactions (or without absence of interactions
for the bond-diluted Ising model).  $Z_{1}$ and $Z_{1}^{\ast }$
are those with a single antiferromagnetic interaction (or absence
of interactions for the bond-diluted Ising model).  The summation
(reduced configurational average for $\tau _{ij}$) is taken over
all realizations with a single antiferromagnetic interaction on
the cluster. Next, we set the coupling constant as $K_{c}+\Delta K$
($\Delta K\ll1$) near the Onsager point. Equation (\ref{eq:lala})
for the slope $\zeta$ thus reduces to
\begin{widetext}
\begin{equation}
\zeta \equiv \frac{1}{T_{c}^{\rm IM}}
\left.\frac{\Delta T}{\Delta p}\right|_{p \to 0} 
=
\frac{Z_{0}^{(s)}(K_{c})}{K_{c}}
\sum \left( \log Z_{1}^{\ast (s)}(K_{c}) - \log Z_{1}^{(s)}(K_{c})\right)
\left(\displaystyle\frac{\mathrm{d}Z_{0}^{\ast(s)}}
{\mathrm{d}K}-\frac{\mathrm{d}Z_{0}^{(s)}}{\mathrm{d}K}\right)^{-1},  
\label{DTc}
\end{equation}
\end{widetext}
where we use the identity $Z_{0}^{\ast}(K_{c})=Z_{0}(K_{c})$ at the
Onsager point [see Eq.~(\ref{Z2})]. To estimate the precise value of
the slope we need to compute the exact value of the partition function
on the finite-size cluster and evaluate the reduced configurational
average over all the possible locations of a single antiferromagnetic
interaction.  The precision of the slope computation depends on the
finite-size cluster used.  This, in turn, depends on the evaluation
of the partition function and the reduced configuration average, all
scaling $\sim {\mathcal O}(N_{B}^{(s)})$ for the hierarchical lattices.
If the calculation converges with an increasing number of bonds we
can estimate the thermodynamic limit value to high precision by a
simple extrapolation.

\subsection{Results}
\label{sec:hie}

\begin{figure}

\includegraphics[width=0.30\columnwidth]{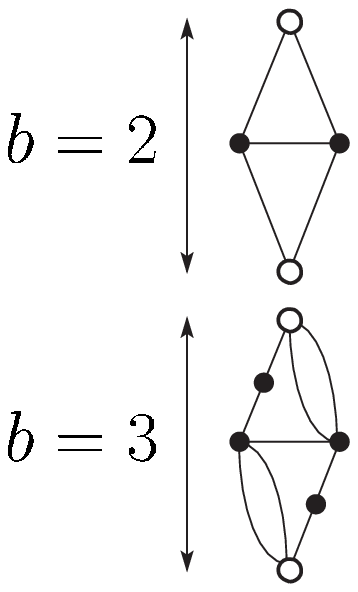}
\hspace*{1.0cm}\includegraphics[width=0.48\columnwidth]{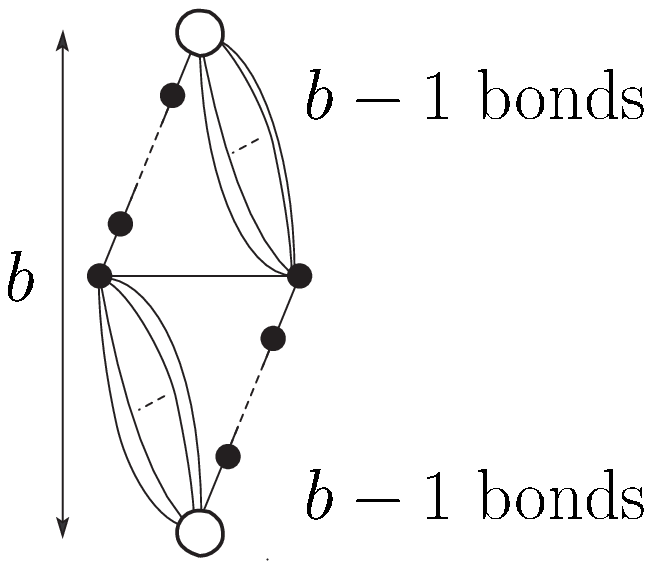}
\caption{
Unit cells of the self-dual hierarchical lattices. 
$b$ is the length of the unit size of the hierarchical lattice.
}
\label{fig:units}
\end{figure}

On the hierarchical lattice (Fig.~\ref{fig:units}) a simple iterative
recursion can be used to estimate the partition function exactly
\cite{berker:79,griffiths:82}.  We can thus obtain an estimate of the
slope by studying sufficiently-large clusters using Eq.~(\ref{DTc}).
At each step of the renormalization an estimate is obtained for a
given cluster size. Results for both bimodal and bond-diluted disorder
are shown in Tables \ref{T1}, \ref{T2}, and \ref{T3}, respectively.
The data clearly converge to a definite value.

\begin{table}[!tb]
\caption{
Results for the phase boundary slope close to the Onsager point
for the (bimodal) $\pm J$ and bond-diluted Ising model on the $b =
2$ self-dual hierarchical lattice.  The bottom line (marked with
`ED') is computed using the simple perturbative approach by Domany
\cite{domany:79,domany:78}.
\label{T1}}
{\footnotesize
\begin{tabular*}{\columnwidth}{@{\extracolsep{\fill}} l r r}
\hline
\hline
$s$   & $\pm J$ 	    & bond diluted         \\
\hline
$1  $ & $3.238670311607941$ & $1.3313280689632047$ \\
$2  $ & $3.213721581962695$ & $1.3295847287696935$ \\
$3  $ & $3.209840157687296$ & $1.3293096628059718$ \\
$4  $ & $3.209227717235686$ & $1.3292661587550924$ \\
$5  $ & $3.209130853824536$ & $1.3292592754393175$ \\
$6  $ & $3.209115527844001$ & $1.3292581862702618$ \\
$7  $ & $3.209113102764629$ & $1.3292580139255745$ \\
$8  $ & $3.209112719032133$ & $1.3292579866545563$ \\
$9  $ & $3.209112658312095$ & $1.3292579823393163$ \\
$10 $ & $3.209112648704037$ & $1.3292579816564926$ \\
$11 $ & $3.209112647183702$ & $1.3292579815484458$ \\
$12 $ & $3.209112646943131$ & $1.3292579815313489$ \\
$13 $ & $3.209112646905064$ & $1.3292579815286436$ \\
$14 $ & $3.209112646899041$ & $1.3292579815282155$ \\
$15 $ & $3.209112646898088$ & $1.3292579815281478$ \\
						   \\
ED    & $3.209112646897908$ & $1.3292579815281351$ \\
\hline
\hline
\end{tabular*}
}
\end{table}

\begin{table}[!tb]
\caption{
Results for the phase boundary slope close to the Onsager point
for the (bimodal) $\pm J$ and bond-diluted Ising model on the $b =
3$ self-dual hierarchical lattice.  The bottom line (marked with
`ED') is computed using the simple perturbative approach by Domany
\cite{domany:79,domany:78}.
\label{T2}}
{\footnotesize
\begin{tabular*}{\columnwidth}{@{\extracolsep{\fill}} l r r}
\hline
\hline
$b=3$ & $\pm J$             & bond diluted         \\
\hline
$1  $ & $3.287620710287408$ & $1.3346666036511970$ \\ 
$2  $ & $3.279280357814356$ & $1.3341234975081161$ \\ 
$3  $ & $3.278702594934052$ & $1.3340857362686508$ \\ 
$4  $ & $3.278662410902251$ & $1.3340831090945090$ \\ 
$5  $ & $3.278659615086040$ & $1.3340829263029411$ \\ 
$6  $ & $3.278659420560328$ & $1.3340829135847418$ \\ 
$7  $ & $3.278659407025692$ & $1.3340829126998396$ \\ 
$8  $ & $3.278659406083984$ & $1.3340829126382702$ \\ 
$9  $ & $3.278659406018462$ & $1.3340829126339864$ \\ 
$10 $ & $3.278659406013903$ & $1.3340829126336883$ \\ 
$11 $ & $3.278659406013586$ & $1.3340829126336676$ \\
						   \\
ED    & $3.209112646897908$ & $1.3292579815281351$ \\
\hline
\hline
\end{tabular*}
}
\end{table}

\begin{table}[!tb]
\caption{
Results for the phase boundary slope close to the Onsager point
for the (bimodal) $\pm J$ and bond-diluted Ising model on the $b =
4$ self-dual hierarchical lattice.  The bottom line (marked with
`ED') is computed using the simple perturbative approach by Domany
\cite{domany:79,domany:78}.
\label{T3}}
{\footnotesize
\begin{tabular*}{\columnwidth}{@{\extracolsep{\fill}} l r r}
\hline
\hline
$s$   & $\pm J$             & bond diluted         \\
\hline
$1  $ & $3.434735249924405$ & $1.3441930333807548$ \\ 
$2  $ & $3.438341429422663$ & $1.3444508638305390$ \\ 
$3  $ & $3.438917613643436$ & $1.3444906297718794$ \\ 
$4  $ & $3.439007297242741$ & $1.3444967410351193$ \\ 
$5  $ & $3.439021128918736$ & $1.3444976788451403$ \\ 
$6  $ & $3.439023254536054$ & $1.3444978226701897$ \\ 
$7  $ & $3.439023580721387$ & $1.3444978447219644$ \\ 
$8  $ & $3.439023630745810$ & $1.3444978481026600$ \\ 
$9  $ & $3.439023638415719$ & $1.3444978486209216$ \\ 
$10 $ & $3.439023639591570$ & $1.3444978487003697$ \\ 
$11 $ & $3.439023639771829$ & $1.3444978487125488$ \\
                                                   \\
ED    & $3.209112646897908$ & $1.3292579815281351$ \\
\hline
\hline
\end{tabular*}
}
\end{table}

For the hierarchical lattice with $b = 2$ we sum up spins up to
$N_{B}^{(s=17)}=5^{17}$.  Equation (\ref{DTc}) for the bimodal Ising
model yields $3.209112646897937\ldots $ for the phase boundary slope,
in agreement (up to 13 digits) with the simple perturbative approach of
Domany [$3.209112646897908\ldots$]. For the bond-diluted Ising model
we obtain for the phase boundary slope $1.3292579815281371\ldots
$ using the duality approach; the simple perturbative approach
of Domany yields $1.3292579815281351\ldots $ (agreement up to 14
digits).  Summarizing, for the $b = 2$ self-dual hierarchical lattices
the duality analysis yields results in perfect agreement with the
simple perturbative approach. 

This implies that the disorder should be marginal or irrelevant on $b=2$
self-dual hierarchical lattices.  However, for $b = 3$ and $b = 4$,
estimates of the phase boundary slope using the duality approach
disagree with the simple perturbative estimate, see Tables \ref{T2}
and \ref{T3}. The simple perturbative result is applicable to all
self-dual lattices if we assume that the effect of the disorder
operator is irrelevant. The fact that for the $b = 3$ and $b=
4$ self-dual hierarchical lattices the perturbative and duality
methods disagree implies that the disorder operator of the effective
Hamiltonian of the bimodal and bond-diluted Ising model is relevant.
Furthermore, this suggests that the phase-boundary slope near the
Onsager point can be used to probe the relevance of disorder effects.

\begin{figure}
\includegraphics[width=\columnwidth]{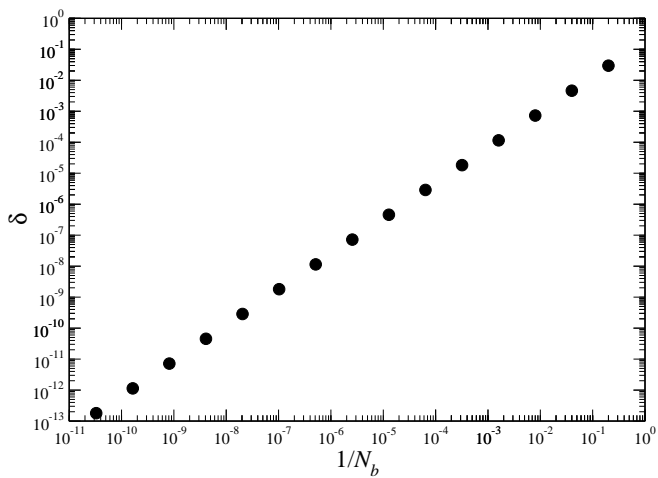}
\caption{
Difference $\delta$ between the estimate for the slope $\zeta$ from
the duality approach [Eq.~(\ref{DTc})] and the expected exact result
\cite{domany:79} as a function of the cluster size $N_B$ for $b = 2$
self-dual hierarchical lattices. The linear behavior in the log-log
plot clearly illustrates the power-law convergence with the system
size.}
\label{fig:diff}
\end{figure}

Finally, our results also show that the data for the slope $\zeta$
converge for an increasing number of bonds to a thermodynamic limiting
value as $\zeta(N_B) = \zeta_\infty + a N_B^{-y}$ with $y \approx
1.15$, see Fig.~\ref{fig:diff} for $b = 2$. The behavior for the
other values of $b$ is qualitatively similar.

\section{Square lattice} 
\label{sec:num}

\subsection{Numerical method}
\label{sec:nums}

Unfortunately, using the duality analysis on square lattices does not
work as well as for hierarchical lattices. Finite-size effects and
apparent systematic deviations prevent from an exact determination
of the phase-boundary slope in this case.  Instead of the duality
analysis, we have conducted numerical simulations to compute a precise
estimate of the slope $\zeta$ on the square lattice for $p \to 1$
to check the value of Domany.

Pfaffian techniques provide access to exact partition
functions in finite spin-glass systems, and developments of these
techniques have become quite sophisticated, allowing {\em exact}
results to be computed for larger system sizes and lower temperatures
than are accessible with more traditional Monte Carlo techniques
\cite{saul:93,galluccio:00,thomas:09}.  Using a dual-lattice
approach \cite{thomas:09,thomas:07}, an algorithm has been recently
developed \cite{thomas:10} to compute exact correlation functions
of the two-dimensional Ising spin glass on the square lattice at all
power-of-two length scales in $\mathcal{O}(N^{3/2})$ time for a system
of $N=L^2$ spins. The correlation function, ${\mathcal C}(\ell)$,
is defined by
\begin{equation}
{\mathcal C}(\ell) = 
\frac{1}{N_\ell}\sum_{i,j \mathrm{~s.t.~} r_{i,j} = \ell} 
\langle S_i S_j\rangle ,
\end{equation} 
with $r_{i,j}$ being the distance between sites $i$ and $j$, $N_\ell
= \sum_{i,j \mathrm{~s.t.~} r_{i,j} = \ell} 1$ is the normalization
factor, and $\langle\cdots\rangle$ represents an average over the canonical
ensemble.
We can then construct a dimensionless ratio---similar to the Binder
cumulant \cite{binder:81}---for each sample
\begin{eqnarray}
b(\ell_1,\ell_2) &=& 
\frac{[{\mathcal C}(\ell_1)]_{\rm av}}{[{\mathcal C}(\ell_2)]_{\rm av}} .
\end{eqnarray}
In the thermodynamic limit the ferromagnetic state has infinite
correlations, i.e., the ratio tends to $b \rightarrow 1$ for any pair
$\ell_1,\ell_2$, while in the paramagnetic state the correlations die
off so that if $\ell_1$ and $\ell_2$ are far apart from one another,
they will tend to $b \rightarrow 0$ (without loss of generality,
assuming that $\ell_1 > \ell_2$).  In order to ensure that $\ell_1$
and $\ell_2$ are far apart, it is useful for them both to scale
with system size $L$.  At the transition temperature $b$ approaches
a constant independent of the system size $L$ (up to corrections
to scaling).  We find empirically that $b(L/2,L/4)$ has very small
finite-size corrections, so we use this ratio for our computations.

To extract $T_c$ from the $b$ crossings, we perform a finite-size
scaling analysis.  It is expected that
\begin{equation} 
\label{eq:scaling}
b \sim \tilde{B}\left(L^{1/\nu}(T-T_c)\right) ,
\end{equation} 
where $\tilde{B}$ is a unknown scaling function.  The values of $T_c$
and $\nu$ responsible for the best scaling collapse are therefore
our best estimates of these quantities.

\begin{figure}
\includegraphics[width=\columnwidth]{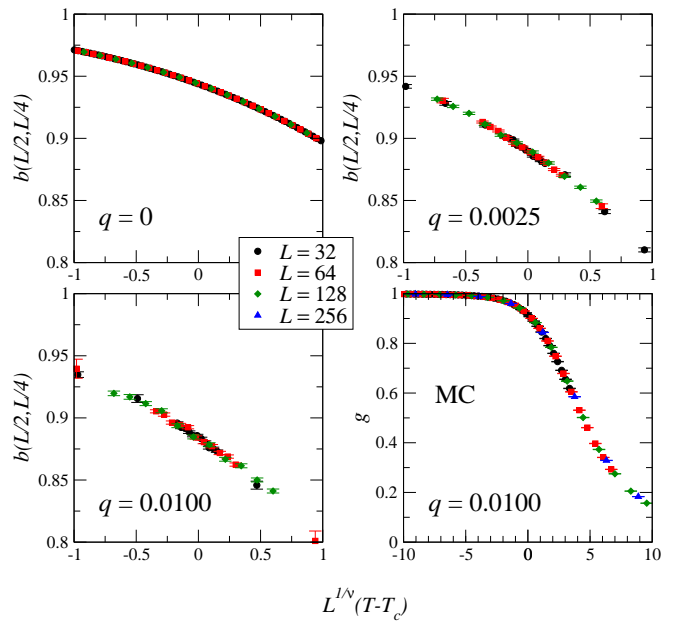}
\caption{
Finite-size scaling collapse for the ratio $b$ as a function of the
scaling variable $L^{1/\nu}(T - T_c)$ for different values of $q$.
Representative numerical data used for this scaling collapse are shown
for $q=0$, $0.0025$, and $0.010$.  No fitting was done for $q=0$:
the numerical result is exact (no disorder averaging is necessary
for $q=0$), so exactly known values for $T_c$ and $\nu$ in the pure
two-dimensional Ising model are used to achieve the data collapse,
therefore illustrating that the method used works.  The panel on the
bottom right labeled with `MC' is a separate finite-size scaling
collapse of the Binder ratio $g$ for independently-computed Monte
Carlo data at $q = 0.0100$ using the value of $T_c$ computed for the
data in the lower left panel.
}
\label{fig:numericalData}
\end{figure}

In order to automatically extract $T_c$ and $\nu$ using standard
curve-fitting techniques, we fit to the functional form of $\tilde{B}$.
We emphasize that the form of $\tilde{B}$ is unimportant except as a
step to automate the procedure: the values we extract for $\nu$ and
$T_c$ give a good scaling collapse which may be verified independently
(as shown in Fig.~\ref{fig:numericalData}), and for this purpose the
form of $\tilde{B}$ may be discarded.  We find that fitting to an
arbitrary (four-parameter) cubic polynomial gives a very good scaling
collapse.  Two independent variables, $L$ and $T$, appear as parameters
in $\tilde{B}$.  We therefore perform a two-dimensional curve fit
using a nonlinear least-squares Levenberg-Marquardt algorithm.
To correctly evaluate the statistical error bars, we perform
a bootstrap analysis. For each value of $p$ and $T$, we have computed $b$
for $~10^4$ different instances of disorder from which we built
a bootstrap sample for the critical parameters \cite{newman:99}.

Letting $q \equiv (1-p)$, simulations have been carried out for
$q=0.0025$ to $0.05$ in increments of $0.025$ up to $N = 128^2$
spins with $10000$ disorder samples.  The results are summarized in
Table \ref{numericalTcTable} and Figs.~\ref{fig:numericalData} and
\ref{fig:numericalPhaseDiagram}.  In Fig.~\ref{fig:numericalData}
representative scaling collapses of the raw data are presented.
For the pure case, $q = 0$, there are no error bars because this is
an exact technique.  Thus the pure result shows the magnitude of the
finite-size effects with this analysis.  A sample result at $q=0.01$
using exchange Monte Carlo is also presented. In this case, the
Binder ratio
\begin{eqnarray}
g&=&\frac{1}{2}
\left(
3 - \frac{[\langle m^4 \rangle]_{\rm av}}{[\langle m^2 \rangle]_{\rm av}^2}
\right)
\end{eqnarray}
is used. Here $\langle \cdots \rangle$ represents an average over Monte
Carlo time. $31$ temperatures in the range $[2.0,2.3]$ are simulated
for systems up to $128^2$ spins. The system is equilibrated for
$2^{16}$ Monte Carlo sweeps; the same time is used for the time average
of the data.  Statistical error bars are determined by performing a
jackknife analysis of $5000$ independent runs.  The Binder ratio is
expected to follow the same scaling relation [Eq.~(\ref{eq:scaling})]
as $b$.  The Monte Carlo data collapse is not achieved by an additional
fit: the values of $T_c$ and $\nu$ extracted for the Pfaffian data are
used, illustrating that that an independent numerical method produces
consistent results, therefore validating our approach using Pfaffian
techniques.  For small amounts of disorder ($q < 0.01$), the computed
values of $\nu$ are consistent with the pure case of the Ising model
where $\nu=1$ \cite{yeomans:92}, and $\nu$ slowly rises to a value of
$\nu\approx 1.10(4)$ for $q=0.05$.  The values of $T_c$ are listed
in Table \ref{numericalTcTable}.  Systematic errors are expected
to come primarily from finite-size effects and therefore should be
much smaller than the statistical errors.  One can then extract a
naive slope from each data point by using a finite difference with
the known exact value at $q=0$. These naive slopes are also shown in
table \ref{numericalTcTable}. This technique does not give a precise
estimate of $\zeta$ (we present a precise estimate from a nonlinear
curve fit below), but we point out that the slope is steeper than
Domany's prediction for every value computed this way.  This is due
to higher-order terms in the small-$q$ expansion.

\subsection{Results}
\label{sec:sqres}

\begin{figure}
\includegraphics[width=\columnwidth]{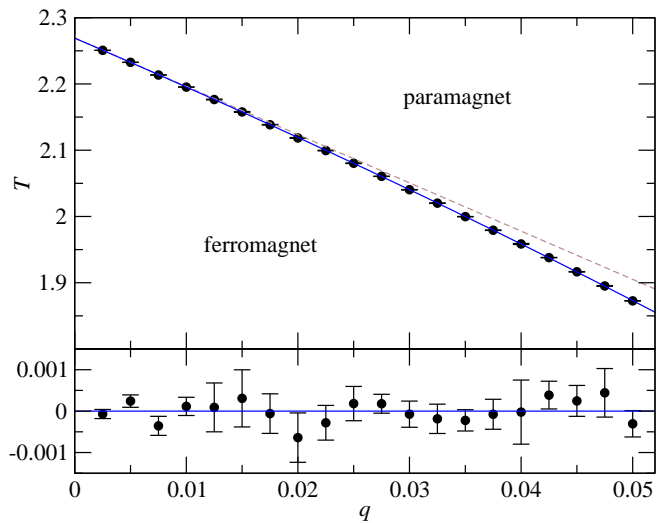}
\caption{
Top: Numerically-computed phase diagram for $q < 0.05$.  The solid
line is the result of a cubic fit to the data.  The dashed line is
the $q \to 0$ limit given by Domany's value for the slope.  Bottom:
The residuals of the fit, computed by subtracting the value of the
cubic fit from each data point, show no clear trends.
}
\label{fig:numericalPhaseDiagram}
\end{figure}

Our results agree with Domany's value, confirming that this expansion
is valid for the two-dimensional Ising model on a square lattice.
In Fig.~\ref{fig:numericalPhaseDiagram} we show the results of a
curve fit to a third-order polynomial for the values of $T_c(q)$.
We find that linear and quadratic curve fits give high chi-square
values and are therefore unlikely to accurately describe the data. We
have also carried out curve fits for third-, fourth- and fifth-order
polynomials: all three give similar results for the slope, and all
three fits have similar chi-square values, implying that the cubic fit
gives correct results.  The value of the cubic fit is the most precise
of these because it has the fewest parameters.  The obtained slope
from the curve fit is $-3.214(17)/T_c(q=0)$, which is consistent with
Domany's value, while the fit value of $T_c(q=0)$ is $2.26906(16)$,
which is consistent with the known exact result.

\begin{table}[!tb]
\caption{ Numerically computed values of $T_c(q)$. In addition, a ``naive''
estimate of the phase boundary slope near the Onsager point is shown where a
linear approximation is performed.
\label{numericalTcTable}}
{\footnotesize
\begin{tabular*}{\columnwidth}{@{\extracolsep{\fill}} l l l}
\hline
\hline
$q$ & $T_c(q)$ & naive slope\\
\hline
0.0025	& 2.25070(11) & 3.258(19) \\
0.0050	& 2.23261(15) & 3.224(13) \\
0.0075	& 2.21349(23) & 3.273(14) \\
0.0100	& 2.19531(22) & 3.256(10) \\
0.0125	& 2.17670(59) & 3.268(21) \\
0.0150	& 2.15779(69) & 3.273(20) \\
0.0175	& 2.13834(48) & 3.295(12) \\
0.0200	& 2.11852(60) & 3.320(13) \\
0.0225	& 2.09947(42) & 3.324(8) \\
0.0250	& 2.08035(41) & 3.329(7) \\
0.0275	& 2.06058(23) & 3.343(4) \\
0.0300	& 2.04037(32) & 3.361(5) \\
0.0325	& 2.02010(35) & 3.378(5) \\
0.0350	& 1.99970(25) & 3.393(3) \\
0.0375	& 1.97927(36) & 3.407(4) \\
0.0400	& 1.95852(77) & 3.423(8) \\
0.0425	& 1.93790(34) & 3.435(3) \\
0.0450	& 1.91649(37) & 3.454(4) \\
0.0475	& 1.89517(59) & 3.470(5) \\
0.0500	& 1.87265(32) & 3.495(3) \\
\hline
\hline
\end{tabular*}
}
\end{table}

\section{Concluding Remarks}

The duality analysis on hierarchical lattices shows that the
value of the slope is not universal for all the self-dual lattices.
This suggests that the simple perturbation theory for the bimodal and
bond-diluted Ising model, without paying attention to the relevance of
the disorder effects, can yield incorrect results.  More generally,
our results illustrate that the effects of the disorder can actually
modify the value of the slope of the paramagnetic--ferromagnetic phase
boundary near the Onsager point for self-dual lattices.  For the $b
= 2$ self-dual hierarchical lattice the value of the slope agrees
with the standard perturbative result by Domany.  If the effects of
the disorder are relevant, the value of the slope changes depending
on the shape of the self-dual lattices sharing the same nonrandom
critical point.  Therefore, the value of the slope can be regarded as
a means for classification of the relevance of the disorder effects
on Ising models defined on self-dual lattices.  We also numerically
reproduce Domany's estimate within statistical error bars
for the square lattice using exact Pfaffian techniques.

Several self-dual models where disorder is a relevant operator have
a close relationship with topologically-protected quantum computation
\cite{kitaev:03,bombin:06}. It is therefore of interest to gain a 
deeper understanding of the effects of disorder on self-dual Ising 
spin systems.

\begin{acknowledgements}

We would like to thank A.~P.~Young and H.~Nishimori for fruitful
discussions.  M.O.~acknowledges support from the Ministry of Education,
Science, Sports and Culture, Grant-in-Aid for Young Scientists (B)
under Grant No.~20740218.  He would also like to thank the Texas A\&M
University Physics and Astronomy Department for their hospitality
during a visit.  M.A.M-D.~and H.B.~acknowledge financial support
from the Spanish MICINN Grant No.~FIS2009-10061, the CAM research
consortium QUITEMAD S2009-ESP-1594, the European Commission PICC:
FP7 2007-2013 (Grant No.~249958), and UCM-BS Grant No.~GICC-910758.
H.G.K.~acknowledges support from the Swiss National Science Foundation
(Grant No.~PP002-114713).  The authors acknowledge Texas A\&M
University for access to their hydra and eos cluster, the Texas
Advanced Computing Center (TACC) at The University of Texas at
Austin for providing HPC resources (Ranger Sun Constellation Linux
Cluster), the Centro de Supercomputaci{\'o}ny Visualizaci{\'o}n de
Madrid (CeSViMa) for access to the magerit cluster, the Barcelona
Supercomputing Center for access to the MareNostrum cluster within
the Spanish Supercomputing Network and ETH Zurich for CPU time on
the Brutus cluster.

\end{acknowledgements}

\bibliography{refs,comments}

\begin{thebibliography}{38}
\bibitem[{com()}]{comment:exceptions}
\bibinfo{note}{There are some rare cases where the Harris criterion fails, such
  as directed polymers \cite{mukherji:95} and spin models \cite{efrat:01} on
  hierarchical lattices.}
\end{thebibliography}

\end{document}